\documentclass[11pt]{article}
\usepackage{epsfig} 
\newcommand{\half}{\mbox{\small{$\frac{1}{2}$}}} 
\newcommand{\MSbar}{\overline{\mbox{MS}}} 

\begin{document}
\title{Crossover exponent in $O(N)$ $\phi^4$ theory at $O(1/N^2)$} 
\author{J.A. Gracey, \\ Theoretical Physics Division, \\ Department 
of Mathematical Sciences, \\ University of Liverpool, \\ Peach Street, \\ 
Liverpool, \\ L69 7ZF, \\ United Kingdom.} 
\date{} 
\maketitle 
\vspace{5cm} 
\noindent 
{\bf Abstract.} The critical exponent $\phi_c$, derived from the anomalous
dimension of the bilinear operator responsible for crossover behaviour in 
$O(N)$ $\phi^4$ theory, is calculated at $O(1/N^2)$ in a large $N$ expansion in 
arbitrary space-time dimension $d$ $=$ $4$ $-$ $2\epsilon$. Its $\epsilon$ 
expansion agrees with the known $O(\epsilon^4)$ perturbative expansion and new 
information on the structure of the five loop exponent is provided. Estimates
of $\phi_c$ and the related crossover exponents $\beta_c$ and $\gamma_c$, using
Pad\'{e}-Borel resummation, are provided for a range of $N$ in three 
dimensions.

\vspace{-16cm}
\hspace{13.5cm} 
{\bf LTH 541} 

\newpage 

Renormalization group techniques have been widely used to study the critical
properties of the scalar quantum field theories underlying a variety of 
condensed matter systems. For instance, Wilson and Fisher, \cite{1,2,3}, 
introduced the technique of extracting numerical estimates of critical 
exponents from the evaluation of the perturbatively calculated renormalization 
group functions at several loop orders in $\phi^4$ theories. Subsequently 
various authors have developed this method to very high loop orders either in 
fixed, (three), spacetime dimensions or in $d$~$=$~$4$~$-$~$2\epsilon$ 
dimensions. The results for various renormalization group functions at,
respectively, six and five loops, which represents the highest orders computed,
are given in \cite{4,5}. In the latter case the results have been extrapolated 
to three dimensions using resummation techniques, \cite{5,6,7}. These exponents
derived by the renormalization group method are competitive with other 
approaches such as the high temperature series expansion and Monte Carlo 
results and are in good agreement with experiment. A recent and comprehensive 
review of the application of the renormalization group in this area is given in
\cite{8}. Recently the critical exponent corresponding to crossover behaviour 
in $O(N)$ $\phi^4$ theory has been calculated to a new degree of accuracy in 
fixed dimension, \cite{9}, where the relevant Feynman diagrams were calculated 
to {\em six} loops. One motivation for that study rests in the realisation that
the $N$ $=$ $5$ theory of superconductivity has been observed in nature, 
\cite{10}. Prior to this the same exponent had been computed to four loops in 
$\MSbar$ in $d$~$=$~$4$~$-$~$2\epsilon$ perturbation theory in \cite{11}, which
built on the lower loop calculations of \cite{3,12}. The resulting numerical 
estimate for that crossover exponent was in agreement with the high temperature
series of \cite{13}. One other field theoretic technique which is used in 
estimating critical exponents is the large $N$ method where the exponents are 
computed order by order in powers of $1/N$. Indeed exploiting the conformal 
properties of the $d$-dimensional Wilson-Fisher fixed point the technique has 
successfully produced the critical exponent $\eta$ at $O(1/N^3)$ in 
$d$-dimensions, \cite{14}, through use of the conformal bootstrap programme. 
Moreover, this method had developed out of the earlier $d$-dimensional critical
point technique of \cite{15,16} which was based on analysing Schwinger Dyson 
equations at criticality in large $N$. In essence the method efficiently 
reproduces the bubble summation which is the main property of $1/N$ expansions 
but more importantly goes beyond the leading order which the conventional 
bubble sum calculations fail to handle easily. Moreover, since the exponents 
are expressed as a function of $d$ they can be expanded in powers of $\epsilon$
and the coefficients compared with those of the same exponents computed in 
conventional perturbation theory. Due to the critical renormalization group the
coefficients must be in agreement. Therefore, the information contained in the 
large $N$ exponents can be exploited, for instance, to gain insight into the 
large order structure of the renormalization group functions at several orders 
in $1/N$. Given the recent interest in the crossover exponent we will focus in 
this letter on its evaluation at a {\em new} order in $1/N$ in $O(N)$ $\phi^4$
theory. Previously the exponent had been calculated at $O(1/N)$ in 
$d$-dimensions in \cite{17,18}. To achieve this we follow the extension of the 
large $N$ fixed point Schwinger Dyson approach of \cite{15,16}, to the 
computation of the anomalous dimensions of composite operators, \cite{19}. 

We recall the essential points of our calculation. The crossover exponent we 
are mainly interested in, $\phi_c$, is computed from the anomalous dimension of
the bilinear traceless symmetric tensor 
\begin{equation}
T^{ab} ~=~ \phi^a \phi^b ~-~ \frac{\delta^{ab}}{N} \phi^c \phi^c
\end{equation} 
where $\phi^a$ is the field of the $O(N)$ $\phi^4$ theory and $1$ $\leq$ $a$
$\leq$ $N$, through the scaling law 
\begin{equation}
\phi_c ~=~ \left( 2 ~-~ \eta_c \right) \nu ~.  
\end{equation} 
This composite operator is relevant for a variety of critical phenomena, 
\cite{9,13,20,21}. The exponent $\nu$ has been computed in $d$-dimensions at 
$O(1/N^2)$ in \cite{16}. To clarify with other work, \cite{9}, the exponent 
$\eta_c$ is related to two other exponents $\eta$ and $\eta_{\cal O}$ by 
\begin{equation}
\eta_c ~=~ \eta ~+~ \eta_{\cal O} 
\end{equation} 
where $\eta$ is the anomalous dimension of the field $\phi^a$ and has been 
computed at $O(1/N^3)$ in \cite{14}. The remaining exponent $\eta_{\cal O}$ is 
the anomalous dimension of the bare composite operator $T^{ab}$ itself. We have
chosen to express the relation for $\phi_c$ in this way since in a gauge theory
the combination $\eta_c$ would be independent of a covariant gauge parameter 
although the analogous $\eta$ and $\eta_{\cal O}$ would each depend on the 
choice of gauge. In the large $N$ critical point method of \cite{19} the 
exponent $\eta_{\cal O}$ is extracted by inserting the operator $T^{ab}$ in a 
two point Green's function and extracting the residue of the simple pole with 
respect to the large $N$ regularization in a well defined fashion according to 
\cite{15,16}. The residues of the simple pole of each Feynman diagram are then 
combined to obtain $\eta_{\cal O}$. Before recalling how this regularization is
introduced we note that the Lagrangian used in the large $N$ technique is  
\begin{equation}
L ~=~ \frac{1}{2} \partial^\mu \phi^a \partial_\mu \phi^a ~+~ \frac{1}{2}
\sigma \phi^a \phi^a ~-~ \frac{3\sigma^2}{2g} 
\label{lag} 
\end{equation} 
where $g$ is the coupling constant and the field $\sigma$ is auxiliary. Its
elimination produces the usual $\phi^4$ interaction. The method of 
\cite{15,16,19} elegantly exploits the properties of the $d$-dimensional 
Wilson-Fisher fixed point in that, for example, the (massless) propagators of 
the fields of (\ref{lag}) have simple power law behaviour. In momentum space, 
representing the propagator by the same letter as the field, the leading 
asymptotic scaling forms in the critical region are  
\begin{equation}
\phi(k) ~ \sim ~ \frac{A}{(k^2)^{\mu-\alpha}} ~~~,~~~ 
\sigma(k) ~ \sim ~ \frac{B}{(k^2)^{\mu-\beta}} 
\label{props} 
\end{equation} 
where $d$ $=$ $2\mu$ and $A$ and $B$ are the momentum independent amplitudes 
which always appear in the combination $z$~$=$~$A^2B$ in the computation of the
Feynman diagrams. The powers of the propagators are related to the usual 
critical exponents by 
\begin{equation} 
\alpha ~=~ \mu ~-~ 1 ~+~ \frac{1}{2} \eta ~~~,~~~
\beta ~=~ 2 ~-~ \eta ~-~ \chi 
\end{equation} 
where $\chi$ is the anomalous dimension of the $\sigma \phi^2$ vertex. It can
also be determined from a scaling law involving $\nu$  
\begin{equation} 
\chi ~=~ \frac{1}{\nu} ~-~ \eta ~-~ 2(\mu-1) 
\end{equation} 
where $\nu$ is proportional to the critical slope of the coupling constant of
the $O(N)$ nonlinear $\sigma$ model which is in the same universality class as
(\ref{lag}) in $2$ $<$ $d$ $<$ $4$. In the large $N$ critical point technique 
the propagators in the Feynman diagrams of a Green's function are represented 
by (\ref{props}).  However, in their present form when they are used to compute
$\eta_{\cal O}$ the leading order large $N$ graphs diverge. To regularize these
infinities the regulator $\Delta$ is introduced by setting 
$\chi$~$\rightarrow$~$\chi$~$+$~$\Delta$. Consequently the Feynman diagrams 
involve poles in $\Delta$ analogous to those in conventional perturbation 
theory where $\epsilon$ in $d$ $=$ $4$ $-$ $2\epsilon$ is the (dimensional) 
regularization. It is the residues of these simple poles in $\Delta$ which are 
then used to extract $\eta_{\cal O}$. It is worth stressing that we will 
compute the exponent in $d$-dimensions where $d$ is arbitrary and $\epsilon$ is
{\em not} used as a regularization. 

For the large $N$ renormalization of the composite operator $T^{ab}$ it turns 
out that only those Feynman diagrams where the operator is not within a 
closed $\phi^a$ field loop will contribute to $\eta_{\cal O}$. This is a 
consequence of the traceless nature of the operator. Diagrams where $T^{ab}$ is
inside a closed $\phi^a$ loop vanish when one computes the group theory factor
of the graph. Therefore, at leading order, $O(1/N)$, there is only one Feynman 
diagram to calculate and applying the method of \cite{19}, we find 
\begin{equation} 
\eta_{{\cal O},1} ~=~ -~ \frac{\mu\eta_1}{(\mu-2)}  
\end{equation} 
where 
\begin{equation} 
\eta_{{\cal O}} ~=~ \sum_{i=1}^\infty \frac{\eta_{{\cal O},i}}{N^i} 
\end{equation} 
with 
\begin{equation} 
\eta_1 ~=~ -~ \frac{4\Gamma(2\mu-2)}{\Gamma(\mu+1)\Gamma(\mu-1)\Gamma(\mu-2)
\Gamma(2-\mu)}  
\end{equation} 
and 
\begin{equation} 
\eta ~=~ \sum_{i=1}^\infty \frac{\eta_i}{N^i} ~.  
\end{equation} 
Consequently, we have 
\begin{equation}
\phi_c ~=~ \frac{1}{(\mu-1)} ~+~ \frac{2\mu\eta_1}{(\mu-1)(\mu-2)N} ~+~ 
O \left( \frac{1}{N^2} \right)  
\label{phi1} 
\end{equation}
which is in {\em exact} agreement with \cite{17,18}, though extracted with a 
minimal amount of effort. In three dimensions, (\ref{phi1}) gives 
\begin{equation} 
\phi_c ~=~ 2 ~-~ \frac{32}{\pi^2 N} ~+~ O \left( \frac{1}{N^2} \right) 
\label{phi13} 
\end{equation} 
or using a Pad\'{e} approximant 
\begin{equation} 
\phi_c ~=~ \frac{2}{\left[1 ~+~ \frac{16}{\pi^2 N} \right]} ~. 
\label{phi13pad} 
\end{equation} 
Interestingly evaluating (\ref{phi13pad}) for $N$ $=$ $2$, $3$, $5$, and $16$ 
we find the respective values $\phi_c$ $=$ $1.105$, $1.298$, $1.510$, and 
$1.816$. These are relatively close to the values obtained by other methods, 
\cite{9}, which are given in Table 1. Indeed the estimate for $N$ $=$ $3$ is 
remarkably good. By contrast the direct evaluation of (\ref{phi13}) gives 
respectively $0.379$, $0.919$, $1.352$ and $1.797$ indicating its poor 
convergence for low $N$.  
{\begin{table}[ht]  
\begin{center} 
\begin{tabular}{|l||l|l|l|} 
\hline
$N$ & $\phi_c$ & $\beta_c$ & $\gamma_c$ \\ 
\hline 
2 & 1.184(12) & 0.830(12) & 0.354(25) \\ 
3 & 1.271(21) & 0.863(21) & 0.41(4) \\ 
4 & 1.35(4) & 0.90(4) & 0.45(8) \\ 
5 & 1.40(4) & 0.90(4) & 0.50(8) \\ 
8 & 1.55(4) & 0.94(4) & 0.61(8) \\  
16 & 1.75(6) & 0.98(6) & 0.77(12) \\ 
\hline
\end{tabular} 
\end{center} 
\begin{center} 
{Table 1. Values of crossover critical exponents from \cite{9}.} 
\end{center} 
\end{table}}  

To determine $\eta_{{\cal O},2}$ we have repeated the method on the $O(1/N^2)$
diagrams. Due to the way the large $N$ expansion orders this would ordinarily
mean that graphs up five loops would have to be calculated. However, when the 
group theory factor is computed only six diagrams remain with a non-zero 
coefficient. These are comprised of four $2$-loop and two $3$-loop graphs. As a
check on our method of calculation we have redetermined $\nu_2$ from the 
evaluation of the exponent $\chi_2$ using the same computer programme written 
in the symbolic manipulation language {\sc Form}, \cite{22}. The method of 
extracting $\chi_2$ is the same as that for $\eta_{{\cal O},2}$ since the 
Feynman diagrams for the latter are equivalent to those for the former when the
operator insertion is replaced by the $\sigma \phi^2$ vertex. Therefore, the 
result of our calculation is 
\begin{eqnarray} 
\eta_{{\cal O},2} &=& \left[ \left( 2\mu ~+~ 5 ~+~ \frac{14}{(\mu-2)} ~+~
\frac{8}{(\mu-2)^2} \right) v^\prime \right. \nonumber \\
&& \left. ~+~ 2\mu ~+~ 2 ~+~ \frac{1}{(\mu-2)} ~-~ \frac{8}{(\mu-2)^2} ~-~ 
\frac{8}{(\mu-2)^3} ~+~ \frac{1}{2(\mu-1)} \right] \eta_1^2  
\end{eqnarray} 
where 
\begin{eqnarray} 
R_1 &=& \psi^\prime(\mu-1) ~-~ \psi^\prime(1) \nonumber \\  
R_2 &=& \psi^\prime(2\mu-3) ~-~ \psi^\prime(2-\mu) ~-~ \psi^\prime(\mu-1) ~+~
\psi^\prime(1) \nonumber \\ 
R_3 &=& \psi(2\mu-3) ~+~ \psi(2-\mu) ~-~ \psi(\mu-1) ~-~ \psi(1) \nonumber \\
v^\prime &=& \psi(2\mu-2) ~+~ \psi(2-\mu) ~-~ \psi(\mu-2) ~-~ \psi(2) 
\end{eqnarray} 
and $\psi(x)$ is defined by $\psi(x)$ $=$ $\Gamma^\prime(x)/\Gamma(x)$ where
$\Gamma(x)$ is the Euler gamma function. Consequently, 
\begin{eqnarray} 
\phi_c &=& \frac{1}{(\mu-1)} ~+~ \frac{2\mu\eta_1}{(\mu-1)(\mu-2)N} 
\nonumber \\
&& +~ \left[ \frac{3\mu^2(8\mu^2-21\mu+14)R_1}{2(\mu-1)(\mu-2)^3} ~-~
\frac{\mu^2(2\mu-3)^2}{(\mu-1)(\mu-2)^3} \left[ R_3^2 ~+~ R_2 \right] \right.
\nonumber \\ 
&& \left. ~~~~~+~ \frac{\mu(4\mu^3-14\mu^2+10\mu+1)}{(\mu-1)^2(\mu-2)^2} 
v^\prime ~-~ 2 ~+~ \frac{6}{(\mu-2)} ~-~ \frac{41}{(\mu-2)^2} \right. 
\nonumber \\
&& \left. ~~~~~-~ \frac{4}{(\mu-2)^3} ~-~ \frac{15}{(\mu-1)} ~+~ 
\frac{1}{(\mu-1)^2} ~+~ \frac{3}{2(\mu-1)^3} \right] \frac{\eta_1^2}{N^2} ~+~ 
O \left( \frac{1}{N^3} \right)  
\label{phi2} 
\end{eqnarray} 
where, for completeness, we note that the values of the other exponents used to 
determine $\phi_c$ are, \cite{15},  
\begin{equation} 
\eta_2 ~=~ \left[ \left( 1 ~-~ \frac{2\mu(\mu-1)}{(\mu-2)} \right) v^\prime ~+~
\frac{(\mu^2+\mu-1)}{2\mu(\mu-1)} ~-~ \frac{2\mu(\mu-1)}{(\mu-2)} ~+~ 
\frac{\mu(3-\mu)}{2(\mu-2)^2} \right] \eta_1^2  
\end{equation} 
and, \cite{16},  
\begin{eqnarray} 
\nu &=& \frac{1}{2(\mu-1)} ~+~ \frac{(2\mu-1)\eta_1}{2(\mu-1)(\mu-2)N} 
\nonumber \\
&& -~ \left[ 3\mu(8\mu-11)R_1 ~-~ \frac{2\mu(2\mu-3)^2}{(\mu-2)} \left[ 
6R_1 ~-~ R_2 ~-~ R_3^2 \right] \right. \nonumber \\ 
&& \left. ~~~~~-~ \frac{2(4\mu^4-12\mu^3+5\mu^2+6\mu-2)}{\mu(\mu-1)} 
v^\prime ~+~ 4\mu^2 ~-~ 2\mu ~+~ 34 ~+~ \frac{8}{(\mu-2)} \right. \nonumber \\
&& \left. ~~~~~+~ \frac{4}{(\mu-1)} ~-~ \frac{3}{(\mu-1)^2} ~+~ 
\frac{2}{\mu^2(\mu-1)} \right] \frac{\mu\eta_1^2}{4(\mu-1)(\mu-2)^2N^2} ~+~ 
O \left( \frac{1}{N^3} \right) \, .  
\end{eqnarray} 
To check the correctness of (\ref{phi2}) we have evaluated $\phi_c$ at 
$O(\epsilon^4)$ in $d$ $=$ $4$ $-$ $2\epsilon$ and compared with the previous
dimensionally regularized four loop $\MSbar$ perturbative calculation of the 
same critical exponent. The result (\ref{phi2}) is in exact agreement which is
a non-trivial check on our computation since only three loop graphs are present
at $O(1/N^2)$. With (\ref{phi2}) we can expand to a {\em new} order in
$\epsilon$ and find  
\begin{eqnarray} 
\phi_c &=& 1 ~+~ \epsilon ~+~ \epsilon^2 ~+~ \epsilon^3 ~+~ \epsilon^4 ~+~ 
\epsilon^5 \nonumber \\
&& -~ \left[ 8\epsilon ~-~ 8\epsilon^3 ~-~ 16( 1 - \zeta(3) )\epsilon^4 ~-~ 
24( 1 - \zeta(4) )\epsilon^5 \right] \frac{1}{N} \nonumber \\ 
&& +~ \left[ 64\epsilon ~-~ 124\epsilon^2 - 4(43 + 60\zeta(3))\epsilon^3 ~+~
(640\zeta(5) - 360\zeta(4) + 976\zeta(3) - 155)\epsilon^4 \right. \nonumber \\
&& \left. ~~~~+~ 2(800\zeta(6) - 1840\zeta(5) + 732\zeta(4) + 128\zeta^2(3) 
- 144\zeta(3) + 61) \epsilon^5 \right] \frac{1}{N^2} ~+~ 
O\left(\frac{\epsilon^6}{N^3} \right) \nonumber \\ 
\end{eqnarray}  
where $\zeta(n)$ is the Riemann zeta function and the order symbol represents
independently higher order terms in $\epsilon$ and $1/N$. The $O(\epsilon^5)$
coefficients will be important in future explicit five loop $\MSbar$
perturbative calculations.

We are now in a position to examine the critical exponents in three dimensions.
For the various ones we are interested in we have 
\begin{eqnarray} 
\phi_c &=& 2 ~-~ \frac{32}{\pi^2 N} ~-~ \frac{64[9\pi^2 + 16]}{9\pi^4N^2} ~+~ 
O \left( \frac{1}{N^3} \right) \nonumber \\ 
\eta_c &=& \frac{32}{3\pi^2 N} ~-~ \frac{512}{27\pi^4N^2} ~+~ 
O \left( \frac{1}{N^3} \right) \nonumber \\ 
\eta_{\cal O} &=& \frac{8}{\pi^2 N} ~+~ O \left( \frac{1}{N^3} \right) ~. 
\end{eqnarray} 
For reference, the other intermediate exponents are 
\begin{eqnarray} 
\eta &=& \frac{8}{3\pi^2 N} ~-~ \frac{512}{27\pi^4N^2} ~+~ 
O \left( \frac{1}{N^3} \right) \nonumber \\ 
\nu &=& 1 ~-~ \frac{32}{3\pi^2 N} ~-~ \frac{32[27\pi^2 + 104]}{27\pi^4N^2} ~+~ 
O \left( \frac{1}{N^3} \right) ~. 
\end{eqnarray} 
In addition we record that the values of two related crossover critical 
exponents are 
\begin{eqnarray} 
\beta_c &=& 1 ~-~ \frac{32[\pi^2 + 8]}{\pi^4N^2} ~+~ 
O \left( \frac{1}{N^3} \right) \nonumber \\ 
\gamma_c &=& 1 ~-~ \frac{32}{\pi^2 N} ~-~ \frac{32[9\pi^2 - 40]}{9\pi^4N^2} ~+~ 
O \left( \frac{1}{N^3} \right) 
\end{eqnarray} 
which are defined through the hyperscaling laws 
\begin{equation}
\beta_c ~=~ 2 \mu \nu ~-~ \phi_c ~~~,~~~ \gamma_c ~=~ 2\phi_c ~-~ 2\mu \nu ~.  
\end{equation} 
Clearly the $O(1/N^2)$ correction to $\phi_c$ is large and the series appears
to diverge. By contrast the $O(1/N^2)$ correction to $\eta_{\cal O}$ vanishes
in three dimensions. We have repeated our earlier Pad\'{e} approach for 
$\phi_c$ to see if the convergence is improved but this does not lead to a
small change to the previous values for the exponents. This is in part due to
the fact that the exponents $\eta$ and $\nu$ do not lend themselves to 
improvement by this approach. Instead one way of gaining estimates from our
large $N$ results is to use the accepted values of $\eta$ and $\nu$ and our
value for $\eta_{\cal O}$. Indeed in \cite{11} the four loop estimate for 
$\phi_c$ was determined in an analogous fashion. Therefore, taking $\eta$ to be
$0.033$ and $0.033$ and $\nu$ to be $0.669$ and $0.705$ for $N$ $=$ $2$ and $3$ 
respectively, \cite{7}, we find the values for $\phi_c$ are $1.044$ and 
$1.196$. These are in poor agreement with the respective results of \cite{9}.  
For the exponents $\beta_c$ and $\gamma_c$ the large $N$ corrections are also 
large for small $N$ and each series appears to converge slowly. To appreciate 
this we have evaluated the above expressions for larger values of $N$. For $N$ 
$=$ $8$ we find $\phi_c$ $=$ $1.475$, $\beta_c$ $=$ $0.908$ and $\gamma_c$ $=$ 
$0.567$. By contrast when $N$ $=$ $16$ our expressions give $1.767$, $0.977$ 
and $0.790$ for the same respective exponents which, by contrast, compare much 
more favourably with the respective values of $1.75(6)$, $0.98(6)$ and 
$0.77(12)$ of \cite{9}. 

In order to improve the convergence of the series we have also examined the 
Pad\'{e}-Borel resummation of the large $N$ series. This involves determining 
the Borel function of the series which is defined by 
\begin{equation} 
\sum_{n=0}^\infty a_n x^n ~=~ \frac{1}{x} \int_0^\infty \! dt ~ e^{-t/x} 
\sum_{n=0}^\infty \frac{a_n t^n}{n!} 
\end{equation} 
and then taking a Pad\'{e} approximant of the integrand given that only several
terms in the series are known. Therefore, for $\phi_c$ its Pad\'{e}-Borel 
estimate is  
\begin{equation} 
\phi_c ~=~ 2N \int_0^\infty \! dt ~ \frac{e^{-Nt}}{\left[ 1 ~-~ a_1 t ~+~ 
(a_1^2 - \half a_2) t^2 \right]} 
\end{equation} 
where 
\begin{equation} 
a_1 ~=~ -~ \frac{16}{\pi^2} ~~~,~~~ 
a_2 ~=~ -~ \frac{32[9\pi^2+16]}{9\pi^4} ~.  
\end{equation} 
We have evaluated the integral numerically for various values of $N$ and 
recorded the results in Table 2 where the estimates for $\beta_c$ and 
$\gamma_c$ by the same method are also given. The final column is the sum of 
the estimates in the second and third columns and represents another way of 
estimating $\phi_c$ through the scaling relation since we have noted that the 
large $N$ series for $\phi_c$ appears to diverge rapidly for low $N$. For $N$ 
$\geq$ $4$ the large $N$ estimates for $\phi_c$ and the sum $\beta_c$ $+$ 
$\gamma_c$ are in fairly reasonable agreement. For $N$ $=$ $2$ and $3$ the 
estimates undershoot those of \cite{9} though the combination $\beta_c$ $+$ 
$\gamma_c$ is closer. For the other exponents the values for $\beta_c$ are 
competitive for $N$~$\geq$~$5$ whilst those for $\gamma_c$ appear to be in good
agreement for the lower range of $N$.
{\begin{table}[ht] 
\begin{center} 
\begin{tabular}{|c||c|c|c|c|} 
\hline
$N$ & $\phi_c$ & $\beta_c$ & $\gamma_c$ & $\beta_c$ $+$ $\gamma_c$ \\ 
\hline 
2 & 0.988 & 0.664 & 0.367 & 1.031 \\ 
3 & 1.187 & 0.768 & 0.459 & 1.227 \\ 
4 & 1.323 & 0.830 & 0.529 & 1.359 \\ 
5 & 1.422 & 0.871 & 0.582 & 1.453 \\ 
8 & 1.603 & 0.934 & 0.689 & 1.623 \\  
16 & 1.790 & 0.980 & 0.817 & 1.797 \\ 
\hline
\end{tabular} 
\end{center} 
\begin{center} 
{Table 2. Pad\'{e}-Borel estimates of crossover exponents.} 
\end{center} 
\end{table}}  

In conclusion, we have provided the $O(1/N^2)$ corrections to a set of 
crossover exponents related to the composite operator $T^{ab}$ in $O(N)$ 
$\phi^4$ theory. Although the leading order exponents could be summed to give
numerical estimates which are competitive with explicit perturbative 
calculations in three dimensions the {\em new} higher order correction 
indicate that the series are slowly converging. Applying the Pad\'{e}-Borel
resummation technique generally improves the estimates in comparison with the
results of \cite{9} though it ought to be borne in mind that $O(1/N^2)$ results
represent only three terms of a series in contrast to \cite{9} which analysed 
six terms of a series. Nevertheless since the critical exponents are computed 
in $d$-dimensions they will complement future higher order perturbative 
calculations and, further, the large $N$ method can equally be applied to the 
determination of crossover exponents of bilinear and other composite operators
to the same large $N$ order in this and other scalar quantum field theories 
which underpin critical phenomena.


\newpage
\begin{thebibliography}{99}
\bibitem{1} K.G. Wilson \& M.E. Fisher, Phys. Rev. Lett. {\bf 28} (1972), 240.
\bibitem{2} K.G. Wilson, Phys. Rev. {\bf B4} (1971), 3174; 
Phys. Rev. {\bf B4} (1971), 3184. 
\bibitem{3} K.G. Wilson, Phys. Rev. Lett. {\bf 28} (1972), 548. 
\bibitem{4} G.A. Baker Jr, B.G. Nickel, M.S. Green \& D.I. Meiron, Phys. Rev.
Lett. {\bf 36} (1977), 1351;  
G.A. Baker Jr, B.G. Nickel \& D.I. Meiron, Phys. Rev. {\bf B17} (1978), 1365; 
S.A. Antonenko \& A.I. Sokolov, Phys. Rev. {\bf E51} (1995), 1894. 
\bibitem{5} H. Kleinert, J. Neu, V. Schulte-Frohlinde, K.G. Chetyrkin \& S.A.
Larin, Phys. Lett. {\bf B256} (1991), 81; 
Phys. Lett. {\bf B319} (1993), 545. 
\bibitem{6} S.G. Gorishny, S.A. Larin \& F.V. Tkachov, Phys. Lett. {\bf A101}
(1984), 120.  
\bibitem{7} J.C. Le Guillou \& J. Zinn-Justin, Phys. Rev. {\bf B21} (1980),
3976.  
\bibitem{8} A. Pelissetto \& E. Vicari, cond-mat/0012164. 
\bibitem{9} P. Calabrese, A. Pelissetto \& E. Vicari, Phys. Rev. {\bf E65} 
(2002), 046115. 
\bibitem{10} S.-C. Zhang, Science {\bf 275} (1997), 1089. 
\bibitem{11} J.E. Kirkham, J. Phys. {\bf A14} (1981), L437. 
\bibitem{12} Y. Yamazaki, Phys. Lett. {\bf A49} (1974), 215. 
\bibitem{13} P. Pfeuty, D. Jasnow \& M.E. Fisher, Phys. Rev. {\bf B10} (1974),
2088.  
\bibitem{14} A.N. Vasil'ev, Yu.M. Pis'mak \& J.R. Honkonen, Theor. Math. 
Phys. {\bf 50} (1982), 127.  
\bibitem{15} A.N. Vasil'ev, Yu.M. Pis'mak \& J.R. Honkonen, Theor. Math. 
Phys. {\bf 46} (1981), 157. 
\bibitem{16} A.N. Vasil'ev, Yu.M. Pis'mak \& J.R. Honkonen, Theor. Math. 
Phys. {\bf 47} (1981), 291.  
\bibitem{17} S. Hikami \& R. Abe, Prog. Theor. Phys. {\bf 52} (1974), 369. 
\bibitem{18} R. Oppermann, Phys. Lett. {\bf A47} (1974), 383. 
\bibitem{19} A.N. Vasil'ev \& M.Yu. Nalimov, Theor. Math. Phys. {\bf 56} 
(1982), 643.
\bibitem{20} M.E. Fisher, Phys. Rev. Lett. {\bf 34} (1975), 1634. 
\bibitem{21} J.M. Kosterlitz, D.R. Nelson \& M.E. Fisher, Phys. Rev. Lett. 
{\bf 33} (1974), 813; 
Phys. Rev. {\bf B13} (1976), 412. 
\bibitem{22} J.A.M. Vermaseren, ``{\sc Form}'' version $2.3$, (CAN, Amsterdam,
(1992). 
\end{thebibliography}
\end{document}